\documentclass[times,twocolumn,final]{elsarticle}
\usepackage{medima}
\usepackage{framed,multirow}

\usepackage{amssymb}
\usepackage{latexsym}

\usepackage{url}
\usepackage{xcolor}

\usepackage{hyperref}

\usepackage{subcaption}
\usepackage{graphicx}
\usepackage{amsmath}
\usepackage{booktabs}

\definecolor{newcolor}{rgb}{.8,.349,.1}

\journal{Medical Image Analysis}

\begin{document}

\verso{Yoav Alon \textit{et~al.}}

\begin{frontmatter}

\title{Neuroplastic graph attention networks for nuclei segmentation in histopathology images}
%\title{Neuroplastic graph attention networks for the generalization of nuclei segmentation in histopathology images across varying experimental configurations}%with residual graph attention networks with learnable graph structure for the generalization of nucleus segmentation.} % \tnoteref{tnote1}}%
%\tnotetext[tnote1]{This is an example for title footnote coding.}

\author[1]{Yoav \snm{Alon}\corref{cor1}}
\ead{ya88@leicester.ac.uk}
%\cortext[cor1]{Corresponding author: 
%  Tel.: +0-000-000-0000;  
%  fax: +0-000-000-0000;}
\author[1]{Huiyu \snm{Zhou}}
\ead{hz143@leicester.ac.uk}
%\author[2]{Given-name4 \snm{Surname4}}

\address[1]{School of Computing and Mathematical Sciences, University of Leicester, University Rd, Leicester LE1 7RH, UK}
%\address[2]{Affiliation 2, Address, City and Postal Code, Country}

\received{January 2022}
%\finalform{10 May 2013}
%\accepted{13 May 2013}
%\availableonline{15 May 2013}
\communicated{Y. Alon and H. Zhou}

\begin{abstract}
  Modern histopathological image analysis relies on the segmentation of cell structures to derive quantitative metrics required in biomedical research and clinical diagnostics. State-of-the-art deep learning approaches predominantly apply convolutional layers in segmentation and are typically highly customized for a specific experimental configuration; often unable to generalize to unknown data.  As the model capacity of classical convolutional layers is limited by a finite set of learned kernels, our approach uses a graph representation of the image and focuses on the node transitions in multiple magnifications. We propose a novel architecture for semantic segmentation of cell nuclei robust to differences in experimental configuration such as staining and variation of cell types. The architecture is comprised of a novel neuroplastic graph attention network based on residual graph attention layers and concurrent optimization of the graph structure representing multiple magnification levels of the histopathological image. The modification of graph structure, which generates the node features by projection, is as important to the architecture as the graph neural network itself. It determines the possible message flow and critical properties to optimize attention, graph structure, and node updates in a balanced magnification loss. In experimental evaluation, our framework outperforms ensembles of state-of-the-art neural networks, with a fraction of the neurons typically required, and sets new standards for the segmentation of new nuclei datasets.

\end{abstract}


\end{frontmatter}

%\linenumbers

\section{Introduction}

While multiple approaches for accurate nucleus segmentation in microscopic image analysis have been developed in the last years, algorithms are usually customized for a specific experimental configuration such as staining, cell type, etc. while largely failing to generalize for other experimental conditions. The 2018 Data Science Bowl \cite{DataScienceBowl} addressed this issue in a public challenge with over 3891 participants providing an annotated dataset joint from multiple experiments with varying conditions. With this attempt to generate a general nucleus segmentation algorithm that can adapt to unknown experiments without the need for human configuration, the benefit for biomedical research is obvious. New unknown cell data as a result of innovation in experimental configuration or novel cell types can be analyzed without specifically customizing and training a neural network thus accelerating biomedical research. 
The vast majority of submitted solutions to the 2018 Data Science Bowl \cite{DataScienceBowl} included deep convolutional neural networks. The best performing solutions featured ensembles supplying multiple networks that produce a combined result. The outstanding approaches of this challenge implemented ensembles of segmentation networks for biomedical image analysis such as fully convolutional networks (FCN) \cite{FCN}, fully convolutional feature pyramid network (FPN) \cite{FPN}, Mask-RCNN \cite{MaskRCNN}, and cell U-Net's \cite{Unet}. Those computing-intensive networks feature a high level of augmentation and post-processing of the images to achieve decent performance but keep short of the performance of specialized networks. 

We argue that state-of-the-art approaches, usually based on encoder-decoder networks, and depending on learnable convolution kernels face a vast model capacity limitation when applied for image analysis tasks that require a generalization to multiple cell types. Where the model capacity of classical convolutional layers is limited by kernels specializing the network to learn very particular features, our approach focuses on node transitions of the graph representation. With the advantage in the ability to learn relational context between pixels of the same and lower magnification. 
Recent advances in adapting transformers \cite{transformers} for image processing tasks in \cite{ImageTransformer} are also based on spatial context and use restriction of self-attention to local neighborhoods to allow higher image sizes to be processed. Similarly, our attention mechanism is restricted to node transitions and can be visualized as edge features. 

Our proposed architecture is composed of a novel network based on residual graph attention layers and a graph representation algorithm that optimizes semantic segmentation for the learned model considering the features assembled in message passing. The modification of graph structure, which generates the node features by projection, is as important to the architecture as the graph neural network itself. It determines the possible message flow and critical properties to optimize attention. This effectively increases the likelihood of identifying features for accurate segmentation. The network structure is learned by modification of node positions, resulting in a feature change representing the underlying pixels. Thus both, learning ideal graph attention convolution and learning an optimal graph structure happen concurrently. Additionally, we weigh the updates by the attention that depends on the positional feature. We devise a balanced magnification focal loss that down-weights simple samples and gives additional weight to hard negatives depending on the positional argument. This allows us to achieve ideal optimization and a high ability to generalize for diverse cell types.   

The main contributions of our research work include: 

\begin{itemize}
  \item Construction of a learnable graph structure that generates the best graph representation for input samples enabling message passing to obtain optimal features for accurate segmentation in multiple levels of magnification. 
  \item Development of a Residual Graph Attention Network architecture based on an advanced self-attention mechanism weighting attention based on relational context. 
  \item Devising a novel loss function that optimizes segmentation for hard examples weighted by magnification features of nodes, thus concurrently optimizing graph structure, attention, and node updates. 
  \item Achieving a high ability to generalize for diverse cell types with a fraction of the neurons required by state-of-the-art networks. 
\end{itemize}

\section{Recent Work}

The motivation for using graph neural networks in cell segmentation is based on multiple factors such as the exhaustion in research using convolutional neural networks, recent progress using image transformers (originally used in natural language processing \cite{transformers} and adapted for images in \cite{ImageTransformer} taking advantage of the spatial context of features), and on the recent advances in efficient optimization of graph neural networks based progress in GPU based frameworks.

\subsection{Cell Segmentation}

Insight into cell mutation and essentials of cell biology can be found at \cite{CellMutation, CellBiology}. In brief, cancer is caused by a genetic mutation that accelerates cell division rates or modifies cell control systems as cycle arrest or cell death. Signals that regularly control cellular growth and death are not responded to adequately. With ongoing cell growth, the deviation from regular cells becomes more drastic. A group of growing cells forms a tumor that may invade neighboring tissues. 

Classical approaches for the application of deep learning including detection and segmentation of cancerous cells in histology are discussed in \cite{MLHisto, MLHisto2, MLHisto3, MLHistoImageAnalysis}. With a wide-ranging survey over historical approaches, the survey provided by \cite{50YearsCellSurvey} gives an insight into segmentation algorithms and the understanding of the cellular mechanism. A comparative survey of cell segmentation algorithms for fluorescent microscopy images is presented by \cite{SegmentationSurveyFluor}. 

Deep learning-based approaches are compared in \cite{NatureDeepLearningCells}. One of the outstanding approaches is the use of a U-Net \cite{UNetCells} that applies a U-Net \cite{Unet}, an encoder-decoder network based on fully convolutional networks (FCN) \cite{FCN}. 

General nuclei segmentation is discussed in \cite{RobustSeg} with an approach based on a computational topology framework. A general study regarding major trends in histopathological cell detection and segmentation was presented by \cite{NucleiSegSurvey} with a thorough overview of various nuclei segmentation algorithms. 

\begin{figure*}
  \centering
  \includegraphics[height=13.5cm]{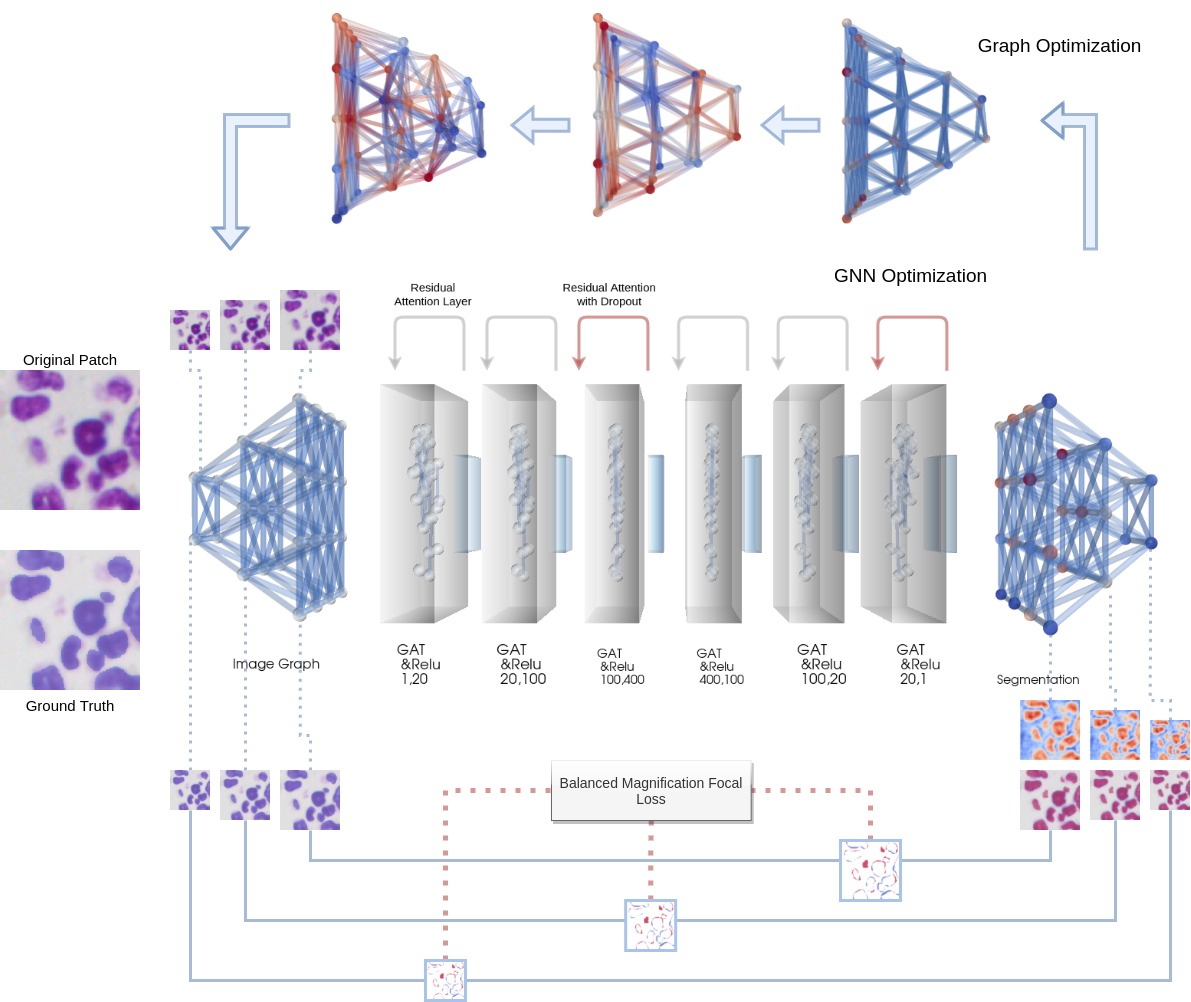}
  \caption{The proposed algorithm features the conversion of a histology image to a multi-magnification graph representation. The extraction of node features depends on node positions relative to the high-resolution layers. At the same time, ground truth patches are extracted in the same manner. In the center of our workflow, a residual graph attention network performs node-wise segmentation that leads to the prediction of segmentation patches followed by the application of a threshold. The layer-wise focal loss is balanced by the magnification and as a result, graph optimization and GNN optimization are achieved. The optimization of the GNN leads to optimal message passing weights and attention coefficients and the optimization of graph structure, achieved by optimization of a position feature, achieves optimization of graph structure that leads to a better choice of relevant magnification. The process of generating graph features from a patch is described in detail in Figure \ref{imgRep}. Preprocessing is not included in this illustration. Best viewed in color.}
  \label{fig:workflow}
\end{figure*}

\begin{figure*}
  \centering
  \begin{subfigure}[b]{0.2\textwidth}
    \includegraphics[height=2.7cm]{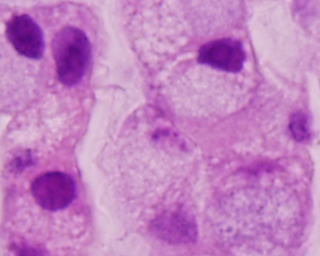}    
    \label{fig:binaryPR}
    \caption{}
  \end{subfigure}
  \begin{subfigure}[b]{0.2\textwidth}
    \includegraphics[height=2.7cm]{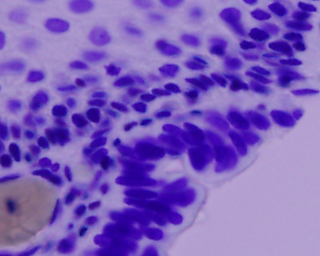}    
    \label{fig:binaryPR}
    \caption{}
  \end{subfigure}
  \begin{subfigure}[b]{0.23\textwidth}
    \includegraphics[height=2.7cm]{images/SmallFluorescent.png}    
    \label{fig:binaryPR}
    \caption{}
  \end{subfigure}
  \begin{subfigure}[b]{0.17\textwidth}
    \includegraphics[height=2.7cm]{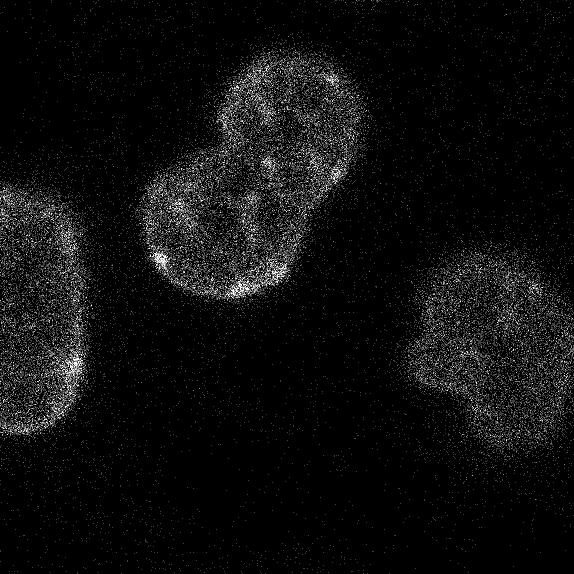}    
    \label{fig:binaryPR}
    \caption{}
  \end{subfigure}
  \begin{subfigure}[b]{0.17\textwidth}
    \includegraphics[height=2.7cm]{images/GrayScale.png}    
    \label{fig:binaryPR}
    \caption{}
  \end{subfigure}
  \caption{Overview of various experimental configurations included in the dataset by staining type (a) pink and purple tissue, (b) purple Tissue, (c) small fluorescent, (d) large fluorescent, and (e) grayscale tissue. Additionally, multiple cell types from human and animal tissue are featured in the dataset. The major challenge of generalization for optimal segmentation of multiple experimental configurations resides in the vast variance of features and cell sizes. Best viewed in color. }
\end{figure*}

\begin{figure*}[t]
  \centering
  \includegraphics[height=5cm]{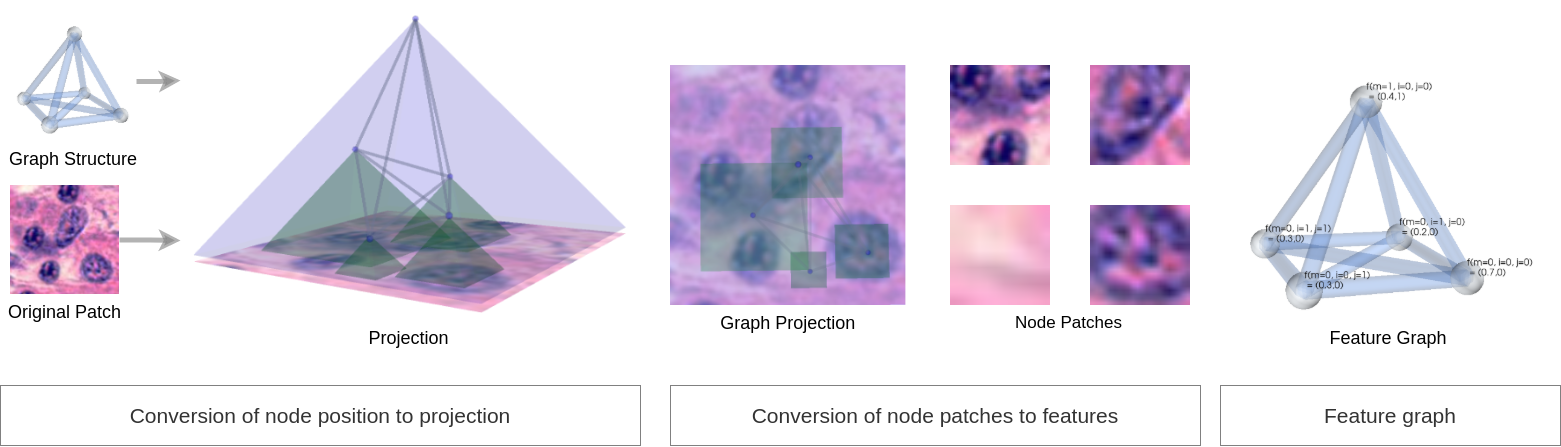}
  \caption{Workflow of generating a graph representation for a given histopathological sample. Initially, the graph structure (here simplified) and original patch are used to extract a projection of segments where equilateral pyramids based on the node positions are constructed. The projection of pixels covered, as described later, is used to construct the feature vector that is then assigned to the respective node. Best viewed in color. }  
  \label{imgRep}
\end{figure*}

\begin{figure*}
  \centering
  \includegraphics[height=7cm]{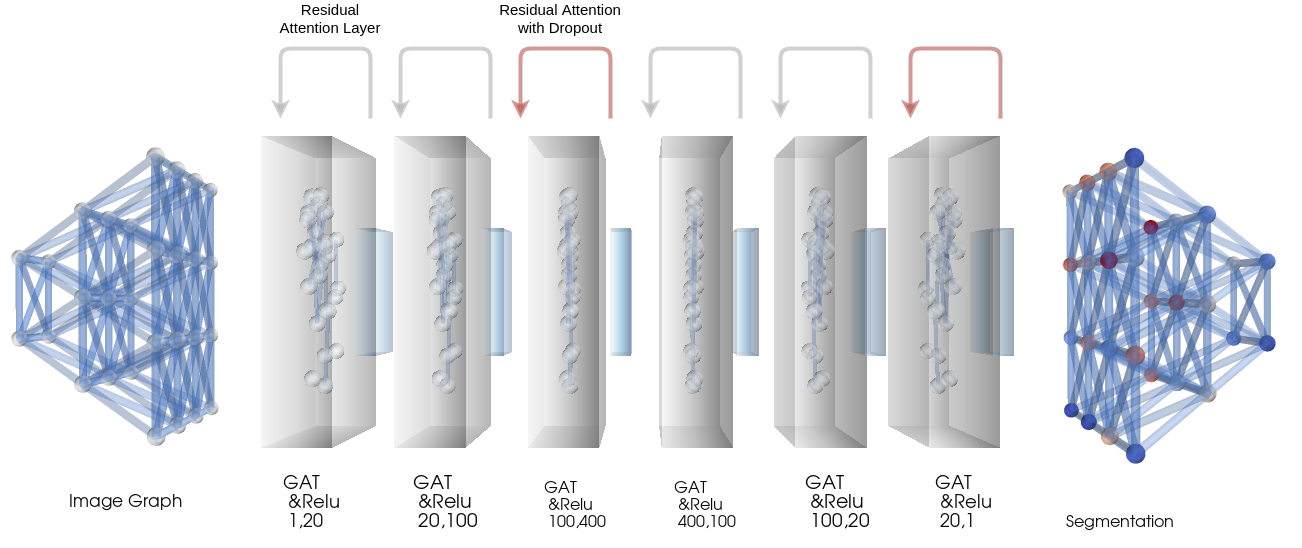}
  \caption{Base network using graph attention convolution layers with dropout and residual connections. The graph structure remains constant where graph convolution performs mechanism of message passing including assembling of neighbor nodes and update of current node feature. The network results in a node-wise segmentation and the extraction of edge-wise attention. The graph is an illustration of a full image graph with multiple magnifications. Best viewed in color. }
  \label{architecture}
\end{figure*}

\subsection{Best approaches of the 2018 Data Science Bowl}

The vast majority of submitted solutions to the 2018 Data Science Bowl \cite{DataScienceBowl} included deep convolutional neural networks. The best performing solutions featured ensembles supplying multiple networks that produce a combined result. The best performing solution by Buslaev, Durnov, and Seferbekov features an ensemble of U-Nets \cite{Unet}. With a high level of augmentation and post-processing and holding 32 large neural networks, the solution was the most demanding solution by computational cost. 
The second-best solution presented by Jiang features a fully convolutional feature pyramid network (FPN) \cite{FPN} where loss is adapted to taking the object size into account and using distance maps rather than binary maps. The third best solution by Lopez-Urrutia implements a single Mask-RCNN \cite{MaskRCNN} where candidate regions are processed. This approach is considered relatively simple in implementation providing very high accuracy in segmentation. The code of these implementations is published on GitHub and more details regarding their implementation can be found in the documentation of \cite{DataScienceBowl}.

\subsection{Image transformers}
Transformers, as originally developed in natural language processing, \cite{transformers} have been introduced to overcome the disadvantages of recurrent neural networks and have been adapted to improve image classification in \cite{ImageTransformer} using learned attention. Restriction of self-attention to local neighborhoods significantly increases the size of images a model can process whereas the size of receptive fields per layer can be larger than in comparable configurations using convolutional neural networks. The bidirectional encoder representation from Image Transformers (BEiT) \cite{BertImageTransformer} allows the further pre-training of image transformers. Similar to vision transformers, the restriction of self-attention to neighborhoods is achieved in our approach using graphs.

\subsection{Graph Neural Networks}
As the graph representation of images in multiple magnifications can best be processed using graph neural networks, we introduce basic concepts such as message passing, aggregation, and updates in graph neural networks. 
For the processing of graphs, permutation invariance is required. Isomorph graphs may appear different but are structurally identical. For that reason adjacency matrices can not be used with feed-forward networks for graph data, that is sensitive to change in node order. Furthermore, the graph structure is non-euclidean, a reason why machine learning around graphs is commonly called geometric deep learning. A representation or embedding of data is learned, where the structure of nodes is conserved and node embeddings are generated. 

%Section seems to be double (also similar in method section)
The process of message passing includes the aggregation of feature vectors of neighbor nodes for the current node. In such a way, we iteratively combine node information for the current node's neighborhood. So, for each graph convolution, a node obtains indirect information of nodes of relation $n$. Most graph neural networks use this process of message passing and only vary in the implementation of functions for update $U$ and aggregation $A$. 

A comprehensive survey of existing approaches for Graph Neural Networks and their application can be found in \cite{GNNs, GNNSurvey2}. The original approach of graph neural networks as presented by Kipf and Welling \cite{GraphConvLayer} used the sum of normalized neighbor embeddings as aggregation in a self-loop. With a Multi-Layer-Perceptron as an aggregator, Zaheer et al. \cite{GraphMLP} preset an approach that propagates states through a trainable MLP. With the development of advanced attention networks, the approach by Velickovic et al. \cite{GraphAttention} focused on attention weights that allow to prioritize the influence of features based on self-learned attention. For heterogeneous graphs with additional edge features Relational Graph Convolution Networks were introduced by Schlichtkrull et. al. \cite{RGCN} to accomplish link prediction and entity classification, enabling the recovery of missing entity attributes for high-dimensional knowledge graphs. 

\section{Proposed Method}

We propose a novel architecture for semantic segmentation of cell nuclei robust to variance in cell types and differences in experimental configuration such as staining. Where the model capacity of classical convolutional layers is limited by kernels specializing the network to learn very specific features, our approach focuses on node transitions. The advantage of this approach is the ability to learn relational context between pixels of the same and lower magnification. Transitions from background to cell body or non-segmentation zone to segmentation zone.

First, we construct a learnable graph structure that generates the best graph representation for input samples enabling message passing to obtain optimal features for accurate segmentation at multiple levels of magnification. 

The main architecture is composed of a novel network based on residual graph attention layers and a graph representation algorithm that optimizes semantic segmentation for the learned model and considers the features assembled in message passing. The modification of graph structure, which generates the node features by projection, is as important to the architecture as the graph neural network itself. It determines the possible message flow and critical properties to optimize attention. Additionally, we weigh the update by the attention that depends on the nodes' positional feature. We devise a balanced magnification focal loss to achieve ideal optimization a high ability to generalize for diverse cell types. 

\subsection{Base graph representation}

The identification of an optimal graph representation for histopathological images has to be considered in the context of the architecture it is processed in. As such the graph structure is an inherent part of the architecture and has a grave impact on segmentation results. 

Here, we describe the generation of the unmodified base graph. Later, the positional arguments of a node are made a learnable parameter and the graph transformation is part of the optimization. Restrictions to modification are discussed in detail in the next section. 

The ability to perform message passing on different levels of magnification allows the model to learn the spatial context of features that improves the overall segmentation performance. The original resolution image provides the base nodes with one node per pixel, with connection to lower magnification layers, which are generated by projection of the node position to the image plane (See Figure \ref{imgRep}).

The number of graph convolutions in the model must allow a message pass that reaches all magnification levels. When optimizing an identical GNN model for two-dimensional graphs, border areas are correctly identified, but since low magnification context is not understood, the cell body is not fully segmented. Consequently, the node features have to reveal information of the node's magnification level. Let a node's feature vector be identified as $f^m_{i,j}$ with $m \in \{0,1, .., n\}$ with $n$ magnification levels and $i,j \in \{0,1, ..., n \}$. As the number of pixels per row and column depend on the magnification level $m$ with $n(m)$. Now the features vector is constructed using projection obtaining the feature $\hat{c}$ and normalized magnification level $\hat{m}$ such that: 
\begin{equation}
  f^m_{i,j} = (\hat{c}, \hat{m})
\end{equation}
The connections allow for direct message passing within one iteration from 
\begin{equation}
  f^m_{i,j} \rightarrow f^m_{k,l}
\end{equation}
with 
\begin{equation}
 d(n^m_{i,j}, n^m_{i,j}) = 1 \text{ and } |m-m'| = 1
\end{equation}
additionally, a king graph that features edges to all neighbors of a node is defined as 
\begin{equation}
  d(n^m_{i,j}, n^m_{i,j}) = 1 \text{ and } |m-m'| = 0
 \end{equation}
Where $d(n_{i,j},n_{k,l} )$ is defined as the minimal node distance between two nodes $n_{i,j}$ and $n_{k,l}$.

To reduce computational costs and allow the network to focus on relevant nodes, the levels of magnification are limited to less than three to five layers for the based graph. The concept of a hierarchical order may resemble classical image convolution that connects features of different resolutions using max-pool layers. But message passing proves more efficient in the use of neurons compared with the limited model capacity by learned kernels in classical convolution. 

\subsection{Graph representation of a histology image}
The workflow of obtaining a graph representation for a given histopathological sample is illustrated in Figure \ref{imgRep}. Initially, the graph structure and original patch are used to extract a projection of segments where equilateral pyramids based on the node positions are constructed. The projection of pixels covered are used to construct the feature vector $f$ that is than assigned to the respective node. In a basic version the extraction of features from a patch can be achieved by averaging the pixel color information such that for each color $c_k$ with $k \in {1,2,3}$: 
\begin{equation}
  f_k = \frac{1}{n}\sum_{i,j} p_{i,j} \text{ and } n = i \dot j
\end{equation}
with $p_{i,j}$ as pixel in position $i,j$ for color $k$. Application of sophisticated perceptrons for feature extractions are ruled out due to a arbitrary number of magnifications.

Several constraints apply in graph mutation, such that the projection must include a minimum number of valid pixels (not including border regions), which restricts nodes' proximity to the ground. As indicated, nodes that are distant from the pyramid's center will also not contain enough pixels. Additionally, node height cannot exceed 1, creating a large image with predominantly empty space. 

\subsection{Mutation of graph structure}

The generation for an optimal graph structure to learn segmentation is a challenge that can be addressed in multiple ways. A deep generative model such as in Li et. al. \cite{DeepGenerativeModel} could learn a graph structure based on an algorithm that generates the graph from scratch. But for image analysis, multiple constraints apply that restrict us to using a constant number of nodes and the constraint that neighbor pixels have to be connected. We, therefore, take the previously described structure $G$ and modify the node positions by applying mutations to the positional feature. In practice, each node is assigned a learnable positional feature that can be optimized. 

So each node $f^m_{i,j}$ is related to a function $\zeta(f^m_{i,j})$ that assigns a three dimensional position $(x,y,z)$: 
\begin{equation}
  \zeta(f^m_{i,j}) = (x,y,z)
\end{equation}
The algorithm for graph modification than performs graph transformation that leaves node-edge connections unchanged but modifies the positional attribute of a graph with the restriction of 
\begin{equation}
\zeta_1(f^m_{i,j}) = (x,y,z) \in [0,..,1] 
\end{equation}
Then the transition of positional function $\zeta$ is performed:
\begin{equation}
\zeta_1(f^m_{i,j}) \rightarrow \zeta_2(f^m_{i,j}) \rightarrow ... \rightarrow \zeta_n(f^m_{i,j})
\end{equation}
The function $\xi$ for modification of $\zeta$ is defined as recursive function 
\begin{equation}
  \zeta_{i+1}(f^m_{i,j}) = \xi(\zeta_{i}(f^m_{i,j}))
\end{equation}
where the nodes $f^m_{i,j}$ remain unchanged for $m = 0$:
\begin{equation}
\xi(\zeta_{i}(f^m_{i,j})) - \zeta_{i}(f^m_{i,j})  = 0 \text{ for } m = 0
\end{equation}
and the node $f^m_{i,j}$ contains the mean of all underlying nodes 
\begin{equation}
 f^{m+1}_{i,j} = \frac{1}{n} \sum f^m_{i,j} \text { where } \delta(f^{m+1}_{i,j}, f^m_{i,j}) < \delta(f^{m+1}_{\hat{i},\hat{j}}, f^m_{i,j})
\end{equation}
Thus, proximity to the next higher magnification node is minimal. 
Now, the function $\xi$ is implemented as using three dense layers with sigmoid activation to optimize node update for a constant number of iterations. Therefore an optimal graph structure that enables learning of node-wise segmentation is achieved. 

\begin{figure*}
  \centering
  \includegraphics[height=15cm]{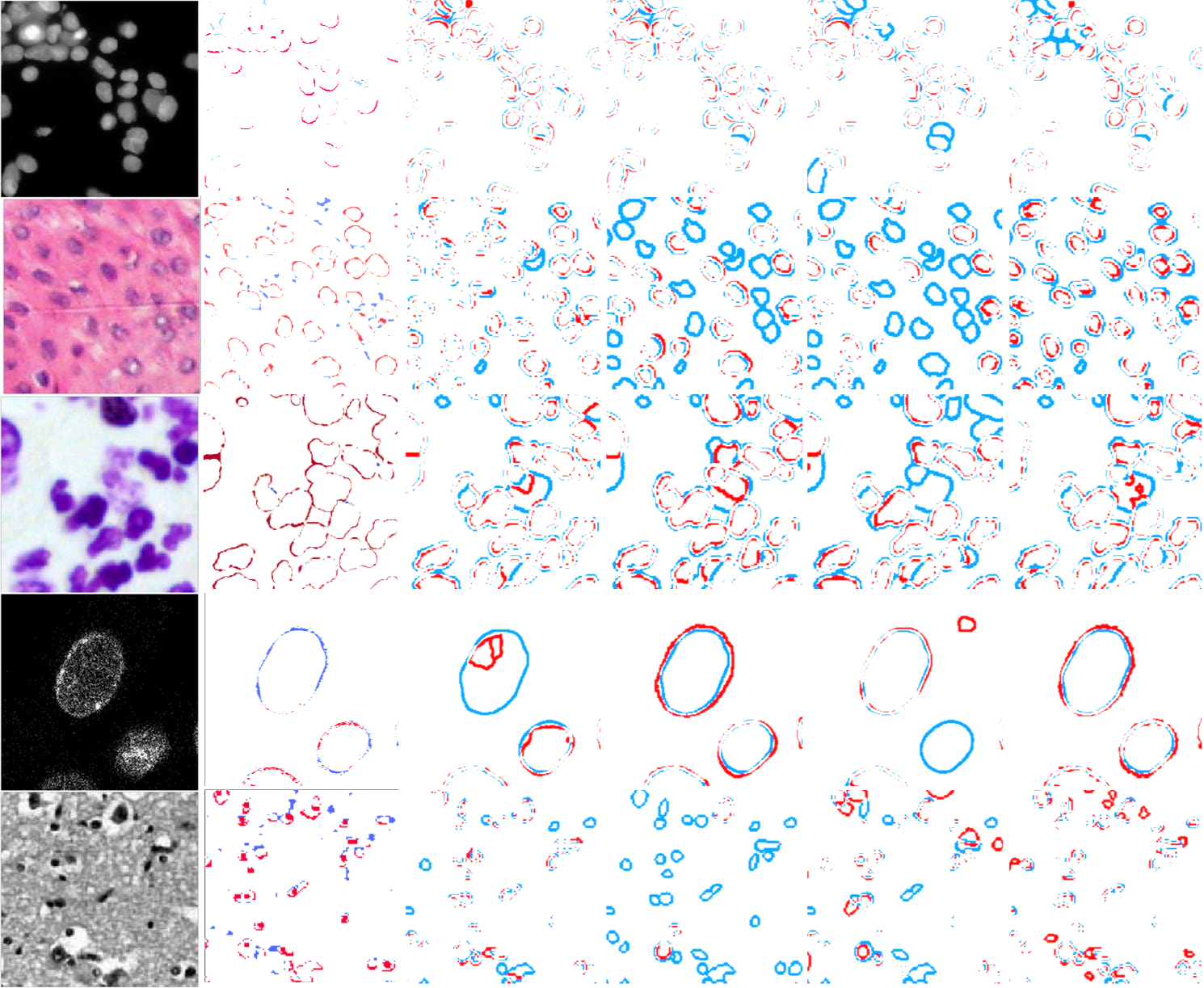}
  \caption{Comparative segmentation maps visualizing pixel-wise alignment compared to annotation. Red areas show extra prediction of cell pixels, where missing prediction is shown in blue. (a) original sample (b) ours (c) U-Net Ensemble (d) FCN, FPN (e) Mask-RCNN (f) Cell profiler. In such a representation an ideal segmentation will reduce the blue and red areas to zero. For medical applications, false-positive errors (type I) or extra annotation will be preferable over false-negative errors (type II) or missing annotation. As confirmed by the experimental metrics the segmentation quality for grayscale tissue is significantly worse due to an unbalanced dataset. The above samples do not necessarily represent the overall metrics for precision and recall calculated for the entire dataset. Best viewed in color. }
  \label{fig:segmentationMaps}
\end{figure*}

\subsection{Aggregation and update}
The process of message passing includes the aggregation of feature vectors of neighbor nodes for the current node. In such a way, we iteratively combine node information for the current node's neighborhood. For each graph convolution, a node obtains indirect information of nodes of relation $n$. Most graph neural networks use this process of message passing and only vary in the implementation of functions for update $U$ and aggregation $A$. In our approach, the edges of each node remain static and the structural modification is achieved by positional modification.

\subsection{Neural Graph Attention based on graph relations}

For the current node: $h_c^{(k)}$ the neighbours $h_i^{(k)}$ with $i \in \{1..n\}$ are aggregated and the current node is updated to $h_1^{(k+1)}$. Every node receives a new embedding. So, for each layer we obtain:
\begin{equation}
    h_c^{(k)}(h_i^{(k)}) \rightarrow h_c^{(k+1)}
\end{equation}
Then, using aggregation function $A$ and update function $U$ we obtain a new node embedding $h_c^{(k+1)}$ for current node $c$ as: 
\begin{equation}
    h_c^{(k+1)} = U^{(k)}(h_u^{k}, A^{(k)}( \{ h_v^{(k)}, \forall v \in N(u)\} ))
\end{equation}
As aggregation function $A$ we use graph convolution networks as proposed by Kipf and Welling in 2016 \cite{GraphConvLayer} where the aggregation is accomplished using: 
\begin{equation}
    h_c^{(k+1)} = \sigma(b^{(k)} + \sum_{j \in N(c)} \frac{1}{c_{ij}} h_j^{(k)}W^{(k)})
\end{equation}
with $c_{ij} = \sqrt{|N(i)|} \sqrt{|N(j)|}$ and $\sigma$ as activation function. Best results are achieved using a rectified linear unit (ReLu) activation function. The Weight $W^{(k)}$ is initialized using Glorot uniform and bias $b$ with zero. Experiments with other aggregator such as Multi-Layer-Perceptron aggregator \cite{GraphMLP} and Graph Attention Networks \cite{GraphAttention} achieve similar results. The Graph Attention Network aggregator can be used to identify additional information regarding nodes that have a high weight in classification. 
Most approaches are based on graphs that hold a local vector representation (one-hot) in each node. Neighborhood aggregation or message passing describes the process of sharing surrounding node embeddings with a reference node. 

With a linear transformation of the lower layer embedding:
\begin{equation}
z_i^{(l)} =W^{(l)}h_i^{(l)}
\end{equation}
a pair-wise un-normalized attention score between two neighbors is calculated:
\begin{equation}
    e_{ij}^{(l)} =\text{LeakyReLU}(\vec a^{(l)^T}(z_i^{(l)}||z_j^{(l)}))
\end{equation}
We then apply softmax to normalize the attention scores for all incoming edges:
\begin{equation}
\alpha_{ij}^{(l)} =\frac{\exp(e_{ij}^{(l)})}{\sum_{k\in \mathcal{N}(i)}^{}\exp(e_{ik}^{(l)})}    
\end{equation}
Then, the neighbour embeddings are aggregated and scaled by the attention scores:
\begin{equation}
    h_i^{(l+1)} =\sigma\left(\sum_{j\in \mathcal{N}(i)} {\alpha^{(l)}_{ij} z^{(l)}_j }\right)
\end{equation}
The general assumption is that attention is agnostic to positional features and thus will not treat nodes of different levels of magnification differently. Since we feature the magnification as part of the node property $f^m_{i,j}$, dense layers can learn the difference between transitions of same level features, and different level features. But the assignment of an optimized node position that samples the mean of underlying node features also contains the linked position $\zeta(f^m_{i,j})$. The relevance for attention has to be learned as a new parameter depending on positional argument, such that $\alpha_{ij}^{(l)}$ should depend on the positional relation $\zeta(f^m_{i,j})$ : 
\begin{equation}
  \alpha_{ij}^{(l)}(\zeta(f^m_{i,j}))
\end{equation}
and we aggregate over 
\begin{equation}
  \sum_{j\in \mathcal{N}(i)} {\alpha_{ij}^{(l)}(\zeta(f^m_{i,j})) z^{(l)}_j }
\end{equation}
So the update becomes:
\begin{equation}
  h_i^{(l+1)} =\sigma\left(\sum_{j\in \mathcal{N}(i)} {\alpha^{(l)}_{ij}(\zeta(f^m_{i,j})) z^{(l)}_j }\right)
\end{equation}
Now similar to the attention itself, the dependency of attention to positional features has to be made learnable. Such that we define transmit position as part of the node feature and extract attention based on $\zeta$.

\subsection{Magnification balanced focal loss for graph nodes}

To effectively optimize graph structure and parameters for weight updating of the previously described network, we need to elaborate on the use of a novel loss function. 

The initial use of a regular binary cross-entropy loss for semantic segmentation results in a good convergence. With the prediction of $\hat{y}$ and the label, or target $y$:
\begin{equation}
  L(y, \hat{y}) = -(y \log (\hat{y}) + (1-y) \log(1- \hat{y}))
\end{equation}
To improve and continue training beyond this limitation focal loss \cite{LossFunctions, FocalLoss}, an extension of cross-entropy down-weights simple samples and give additional weight to hard negatives using: 
\begin{equation}
FL(p_t) = - \alpha_t (1 - p_t)^{\gamma} log(p_t)
\end{equation}
with $\gamma, \alpha \in [0,1]$ and the modulation fact $(1-p_t)^{\gamma}$.
and $p_t = p$ for $y = 1$ and $p_t = 1-p$ otherwise.

Decent results can be achieved using focal loss and binary cross-entropy for the highest magnification level nodes $z_0$ and ignoring the lower magnification nodes. To exploit contextual information of multiple layers of magnification, all nodes must influence the loss. Otherwise transitions from feature to $f(z_i) \rightarrow f(z_{i+1})$ might create virtual feature vectors that tend to slow down optimization. 

The loss of a node is scaled by a three dimensional scalar field $\zeta(n_{x,y,z})$ attributing importance to the positional feature of a node. 
\begin{equation}
  \begin{split}
  \zeta_x(n_{x,y,z}) = \cos(\pi \cdot x) \cdot \sin(\pi \cdot y) \\
  \zeta_y(n_{x,y,z}) = \sin(\pi \cdot x) \cdot \sin(\pi \cdot y) \\
  \zeta_z(n_{x,y,z}) = z \cdot \cos(\pi \cdot y)
  \end{split}
\end{equation}

Thus less attention is attributed to nodes close outside a basis pyramid. 
And the balanced focal loss can be calculated with: 
\begin{equation}
  FL_{MB}(p_t) = - \alpha_t \cdot \zeta(n_{x,y,z}) \cdot (1 - p_t)^{\gamma} log(p_t) 
\end{equation}
%The optimization constraints structural modification preventing nodes being close to constraints and at the same time a higher weight is given to higher magnifications and poses that don't an extensive amount ofh background. 
The optimization constraints for structural modification prevent node proximity and positional features that lead to a mostly empty projection patch. 

\section{Experiments}

\subsection{Technical specifications}
The experiments were performed using GPUs on Alice2, part of the High-performance computing (HPC) Cluster of the University of Leicester. Experiments performed locally use an Nvidia Tesla 10 8GB. For the implementation Pytorch 1.7 on Python 3.6 with the graph library DGL was used. The codebase will be published upon publication of this manuscript. 

\subsection{Implementation}
The workflow includes preprocessing of images and graph generation based on image patches. Then graph batches are provided and graph convolution is performed on the graphs, modifying the node feature vectors that hold information of position and segmentation. Preprocessing includes various steps of image augmentation such as modification of brightness etc. After optimization of graph structure and segmentation, the highest level nodes are reconstructed to a full image. 

\begin{figure*}
  \centering
  \begin{subfigure}[b]{0.15\textwidth}
    \caption*{Original}
    \includegraphics[height=2.7cm]{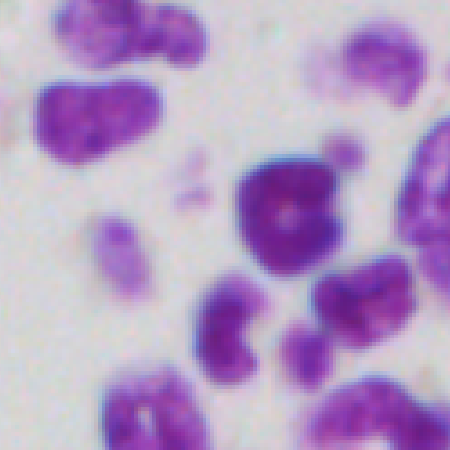}        
    %\caption{}
  \end{subfigure}
  \begin{subfigure}[b]{0.15\textwidth}
    \caption*{Ground Truth}
    \includegraphics[height=2.7cm]{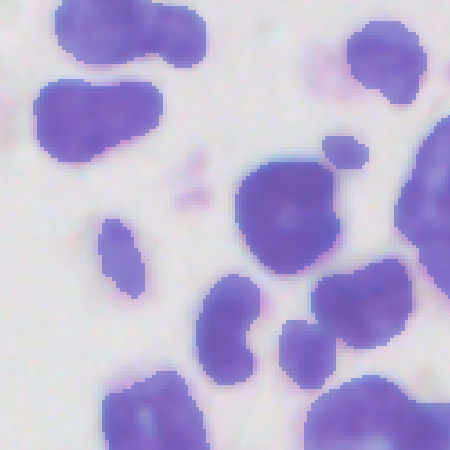}        
    %\caption{}
  \end{subfigure}
  \begin{subfigure}[b]{0.15\textwidth}
    \caption*{Prediction TH}
    \includegraphics[height=2.7cm]{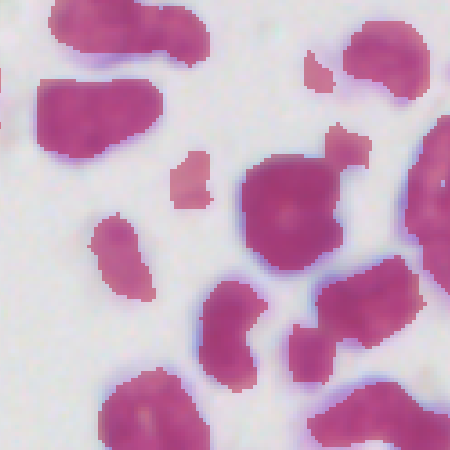}        
    %\caption{}
  \end{subfigure}
  \begin{subfigure}[b]{0.15\textwidth}
    \caption*{Prediction raw}
    \includegraphics[height=2.7cm]{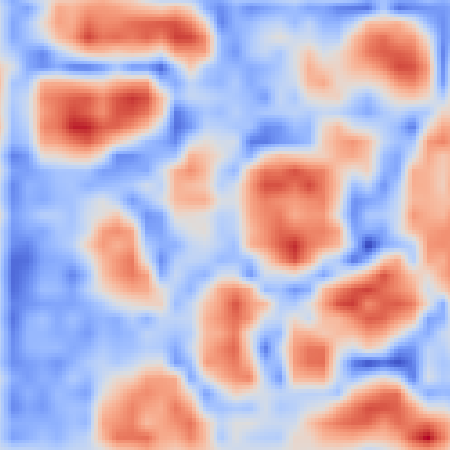}        
    %\caption{}
  \end{subfigure}
  \begin{subfigure}[b]{0.15\textwidth}
    \caption*{Overlay}
    \includegraphics[height=2.7cm]{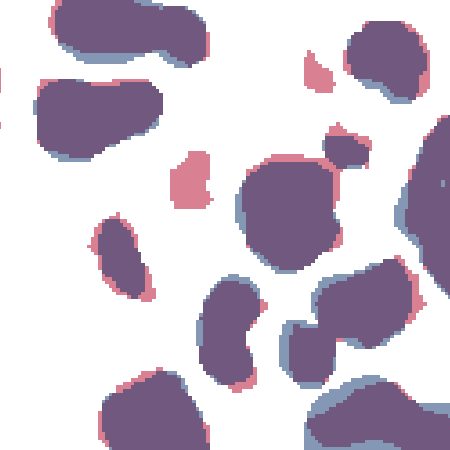}        
    %\caption{}
  \end{subfigure}
  \begin{subfigure}[b]{0.15\textwidth}
    \caption*{Difference}
    \includegraphics[height=2.7cm]{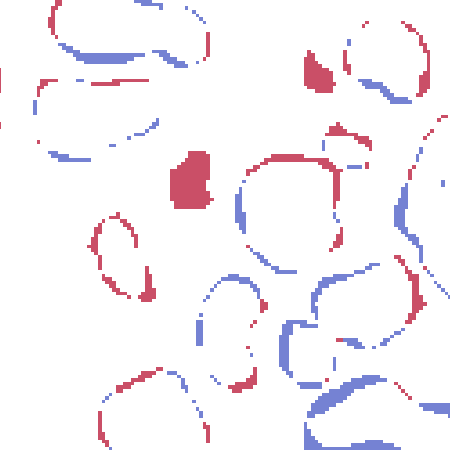}        
    %\caption{}
  \end{subfigure}
  \raisebox{1.6cm}{\rotatebox[origin=t]{90}{pink tissue}}

  \begin{subfigure}[b]{0.15\textwidth}
    \includegraphics[height=2.7cm]{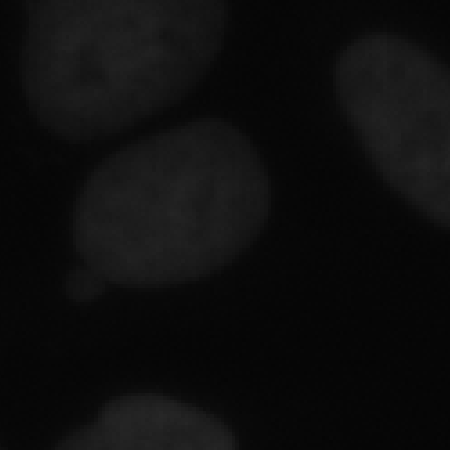}        
    %\caption{}
  \end{subfigure}
  \begin{subfigure}[b]{0.15\textwidth}
    \includegraphics[height=2.7cm]{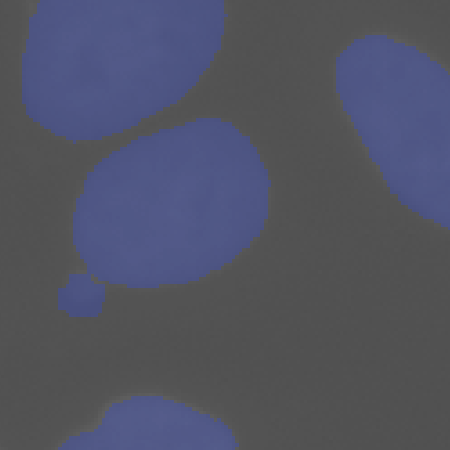}        
    %\caption{}
  \end{subfigure}
  \begin{subfigure}[b]{0.15\textwidth}
    \includegraphics[height=2.7cm]{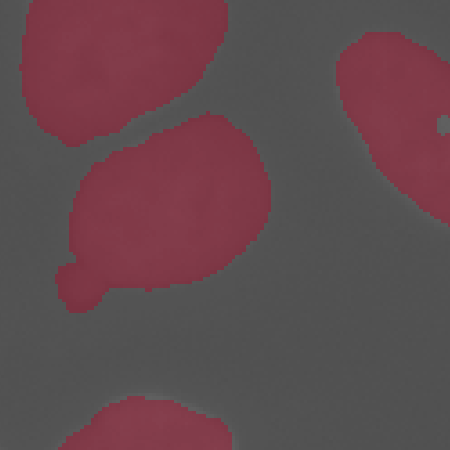}        
    %\caption{}
  \end{subfigure}
  \begin{subfigure}[b]{0.15\textwidth}
    \includegraphics[height=2.7cm]{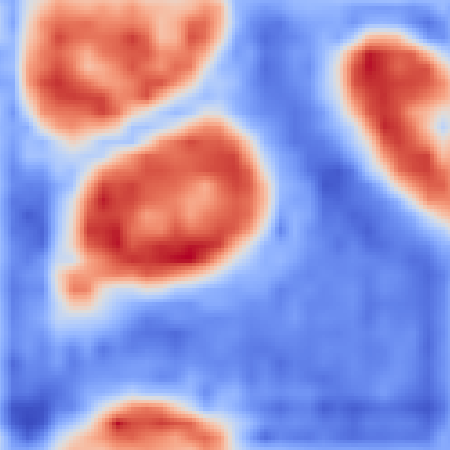}        
    %\caption{}
  \end{subfigure}
  \begin{subfigure}[b]{0.15\textwidth}
    \includegraphics[height=2.7cm]{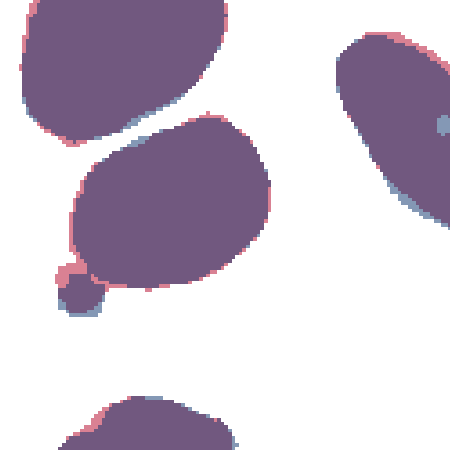}        
    %\caption{}
  \end{subfigure}
  \begin{subfigure}[b]{0.15\textwidth}
    \includegraphics[height=2.7cm]{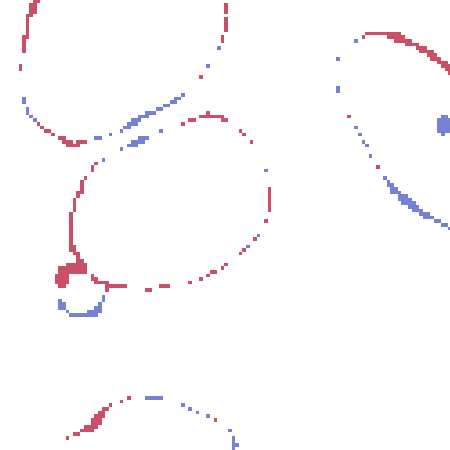}        
    %\caption{}
  \end{subfigure}  
  \raisebox{1.6cm}{\rotatebox[origin=t]{90}{large fluorescent}}

  \begin{subfigure}[b]{0.15\textwidth}
    \includegraphics[height=2.7cm]{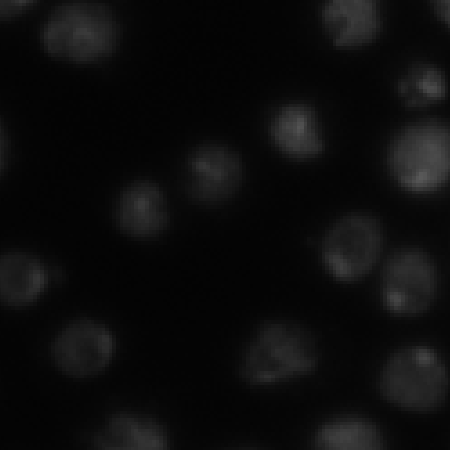}        
    %\caption{}
  \end{subfigure}
  \begin{subfigure}[b]{0.15\textwidth}
    \includegraphics[height=2.7cm]{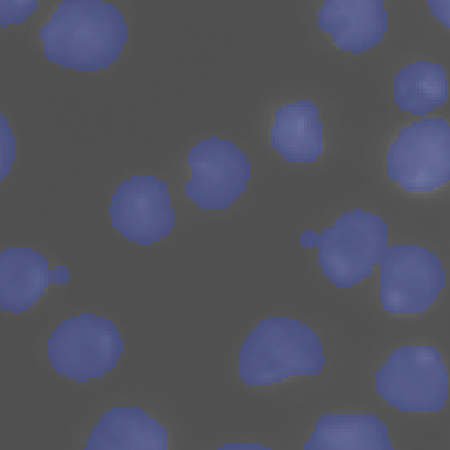}        
    %\caption{}
  \end{subfigure}
  \begin{subfigure}[b]{0.15\textwidth}
    \includegraphics[height=2.7cm]{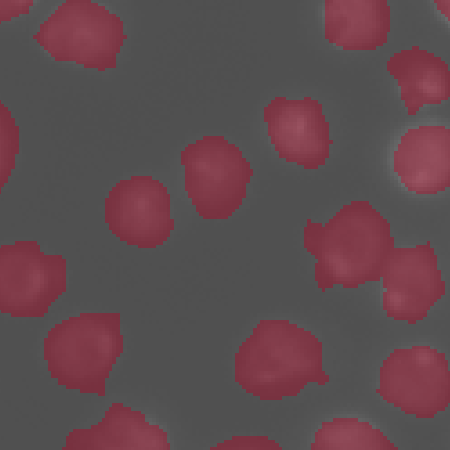}        
    %\caption{}
  \end{subfigure}
  \begin{subfigure}[b]{0.15\textwidth}
    \includegraphics[height=2.7cm]{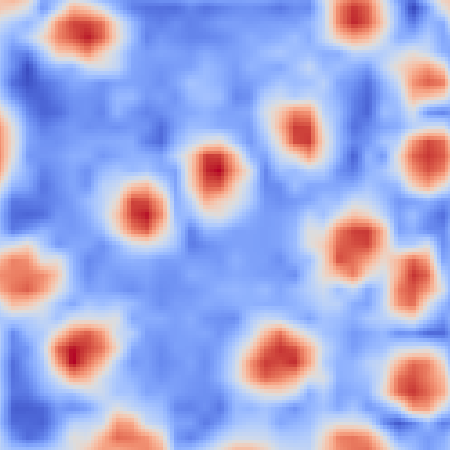}        
    %\caption{}
  \end{subfigure}
  \begin{subfigure}[b]{0.15\textwidth}
    \includegraphics[height=2.7cm]{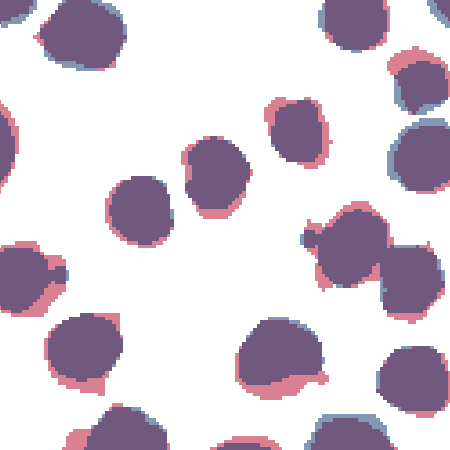}        
    %\caption{}
  \end{subfigure}
  \begin{subfigure}[b]{0.15\textwidth}
    \includegraphics[height=2.7cm]{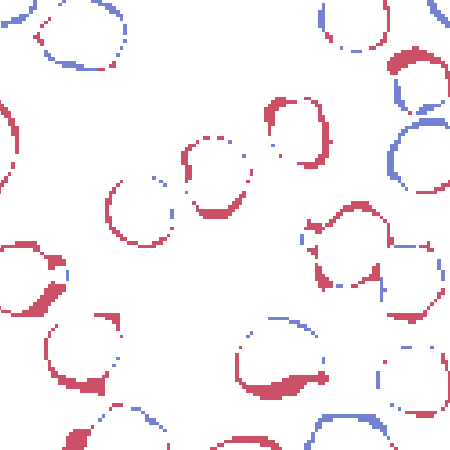}        
    %\caption{}
  \end{subfigure}    
  \raisebox{1.6cm}{\rotatebox[origin=t]{90}{small fluorescent}}

  \caption{Cell samples with their ground truth, threshed and raw prediction, overlay, and ground truth-prediction difference. The raw prediction consists of the weighted reconstruction of multiple magnifications. The threshed prediction is the final network output. The prediction-ground-truth difference allows insight into the networks' performance in terms of extra or missing prediction. Best viewed in color. }
\end{figure*}

\subsection{Dataset}
The dataset is provided by the nuclei segmentation data of the 2018 Data Science Bowl \cite{DataScienceBowl}. The dataset contains annotated histopathological nuclei images based on a variety of different cell types, staining types, magnifications, and experimental variations. The challenge aimed to select the best approaches to generalize segmentation for various experimental configurations without the need for configuration or human interaction. An overview of the best-performing algorithms is given in the recent work section of this work. Multiple laboratories contributed a total of 841 images with a wide biological and experimental variability from 30 biological experiments. The first stage training set includes 670 images with 65 evaluation images, and 106 images for the second stage evaluation thus allowing for realistic evaluation of a general algorithm for nuclei segmentation of unknown images. The annotation was performed by expert biologists manually delineating cells. A total of 37,333 cells were annotated.

\begin{figure}
  \centering
  \includegraphics[height=8cm]{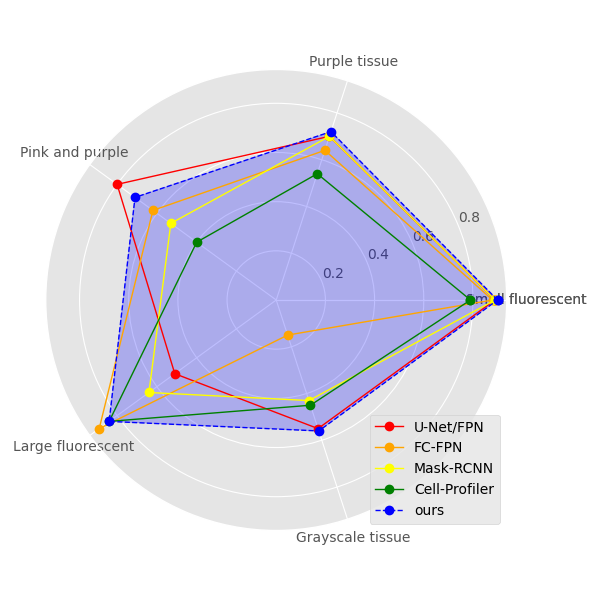}
  \caption{Performance by experimental variation type. The ability to generalize well for various cell types can be evaluated when the comparative F1 score at 0.7 IoU is calculated for each cell class. Comparison with best-performing methodologies from the 2018 data science owl \cite{DataScienceBowl}. Our model covers a larger area than other models indicating a higher ability to generalize for the observed cell types. Best viewed in color.}
  \label{radarPlot}
\end{figure}

\begin{figure*}
  \centering
  \begin{subfigure}[b]{0.48\textwidth}
    \includegraphics[height=6.3cm]{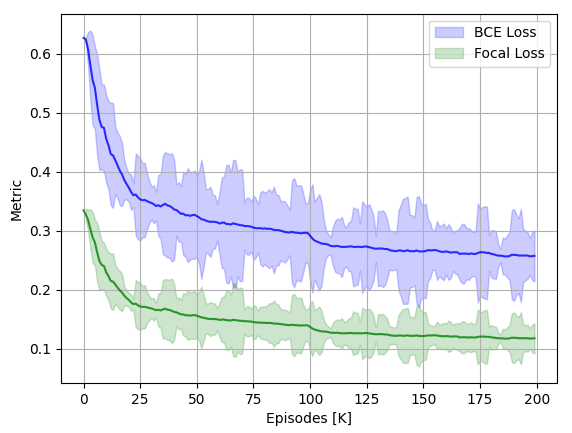}
    \label{loss}
    \caption{}
  \end{subfigure}
  \begin{subfigure}[b]{0.48\textwidth}
    \includegraphics[height=6.5cm]{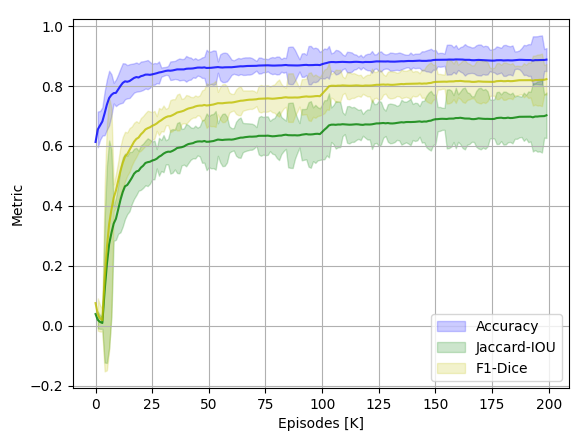}
    \label{fig:metrics}
    \caption{}
  \end{subfigure}
  \caption{Loss per 1000 episodes (a) specifying binary cross-entropy loss (blue) and magnification balanced focal loss (green). Magnification balanced focal loss is based on a focal loss that is weighted higher for high levels of resolution. As an extension of cross-entropy focal loss down-weights simple samples and gives additional weight to hard negatives. Test metrics per 1000 episodes detailing Accuracy, Jaccard-IOU and F1-Dice score. Best viewed in color. }
\end{figure*}

\subsection{Metrics}
We follow the evaluation metrics of the 2018 Data Science Bowl \cite{DataScienceBowl}. With precision $P(t)$ and recall $R(t)$ of: 
\begin{equation}
P(t) = \frac{TP(t)}{TP(t) + FP(t)} \text{ and } R(t) = \frac{TP(t)}{TP(t) + FN(t)}
\end{equation}
Thus using the notion of true positives (TP), etc. For the official competition score S \cite{DataScienceBowl} multiple IoU thresholds were used having: 
\begin{equation}
S = \frac{1}{|T|} \sum_{t \in T} \frac{TP(t)}{TP(t) + FP(t) + FN(t)} 
\end{equation}
with T = \{0.1,0.15,...,0.95\}. The metrics used for segmentation evaluation are Jaccard loss based on intersection over union and the related Dice-Loss (F1). As previously discussed, both metrics can alternatively be adapted as loss functions for the training of the network.  
\begin{equation}
  IOU = \frac{\sum y \hat{p} + 1}{\sum y + \sum \hat{y} + 1}
\end{equation}
And F1-Score (Dice) :
\begin{equation}  
  F1 = \frac{2 \sum y \hat{y}}{\sum y^2 + \sum \hat{y}^2 + \epsilon}
\end{equation}
The experiments required adaption various hyper-parameters, such as learning rate, to achieve optimal training. 

\begin{table*} 
  \centering 
  \begin{tabular}{l c c c c c c c c c} 
  & \multicolumn{3}{c}{\textbf{Scores}}& & \multicolumn{1}{c}{\textbf{Alignments}} \\ 
  \cmidrule(l){3-4} 
  \cmidrule(l){5-7} 
  \textbf{Approach} & Core Model & Competition Score & Average F1 & Recall at 0.7 IoU & Missed at 0.7 IoU & Extra at 0.7 IoU\\ 
  \midrule     
  topcoders & U-Net/FPN & 0.6316 & 0.7120 & 0.7762 & 0.2238 & 0.1455\\
  Jacobkie & FC-FPN & 0.6147 & 0.6987 & 0.6914 & 0.3086 & 0.1504\\
  Deep Retina & Mask-RCNN & 0.6141 & 0.7008 & 0.6807 & 0.3193 & 0.1090\\
  CellProfiler & var & 0.5281 & 0.6280 & 0.5935 & 0.4065 & 0.3955\\
  \midrule
  \textbf{ours} & GAT & 0.6762 & 0.7384 & 0.7931 & 0.2069 & 0.1425\\
  \bottomrule 
  \end{tabular}
  \caption{Experimental comparison with best-performing methodologies from the 2018 data science bowl \cite{DataScienceBowl}. The comparative metrics are detailed in the experimental section of this work. As mentioned by the competition organizers, the F1 score is closely correlated to the competition score. For various IoU thresholds between target and predicted segmentation, F1 scores were computed and averaged.} 
  \label{tab:mainComparison} 
  \end{table*}

  \begin{table*} 
    \centering 
    \begin{tabular}{l c c c c c c c c} 
    & \multicolumn{5}{c}{\textbf{F1 score at 0.7 IoU}}&  \\ 
    \cmidrule(l){2-6} 
    %\cmidrule(l){5-6} 
    \textbf{Approach} & Small fluorescent & Purple tissue & Pink and purple & Large fluorescent & Grayscale tissue\\ 
    \midrule     
    topcoders & 0.89 & 0.70 & 0.80 & 0.51 & 0.55\\
    Jacobkie  & 0.88 & 0.64 & 0.62 & 0.89 & 0.15\\
    Deep Retina  & 0.89 & 0.70 & 0.53 & 0.64 & 0.43\\
    CellProfiler & 0.79 & 0.54 & 0.40 & 0.84 & 0.45\\
    \midrule
    \textbf{ours}  & 0.90 & 0.72 & 0.71 & 0.84 & 0.56\\  %calculate averages as well !!!! should be 3.55 in sum
    \bottomrule 
    \end{tabular}
    \caption{Performance by experimental variation type. Comparison with best-performing methodologies from the 2018 data science bowl \cite{DataScienceBowl}. The average F1 score at 0.7 IoU by experimental configuration type. The comparative metrics are detailed in the experimental section of this work. } 
    \label{tab:ExperimentalVariation} 
    \end{table*}

    \begin{figure*}
      \centering
      \begin{subfigure}[b]{0.45\textwidth}
        \includegraphics[height=8cm]{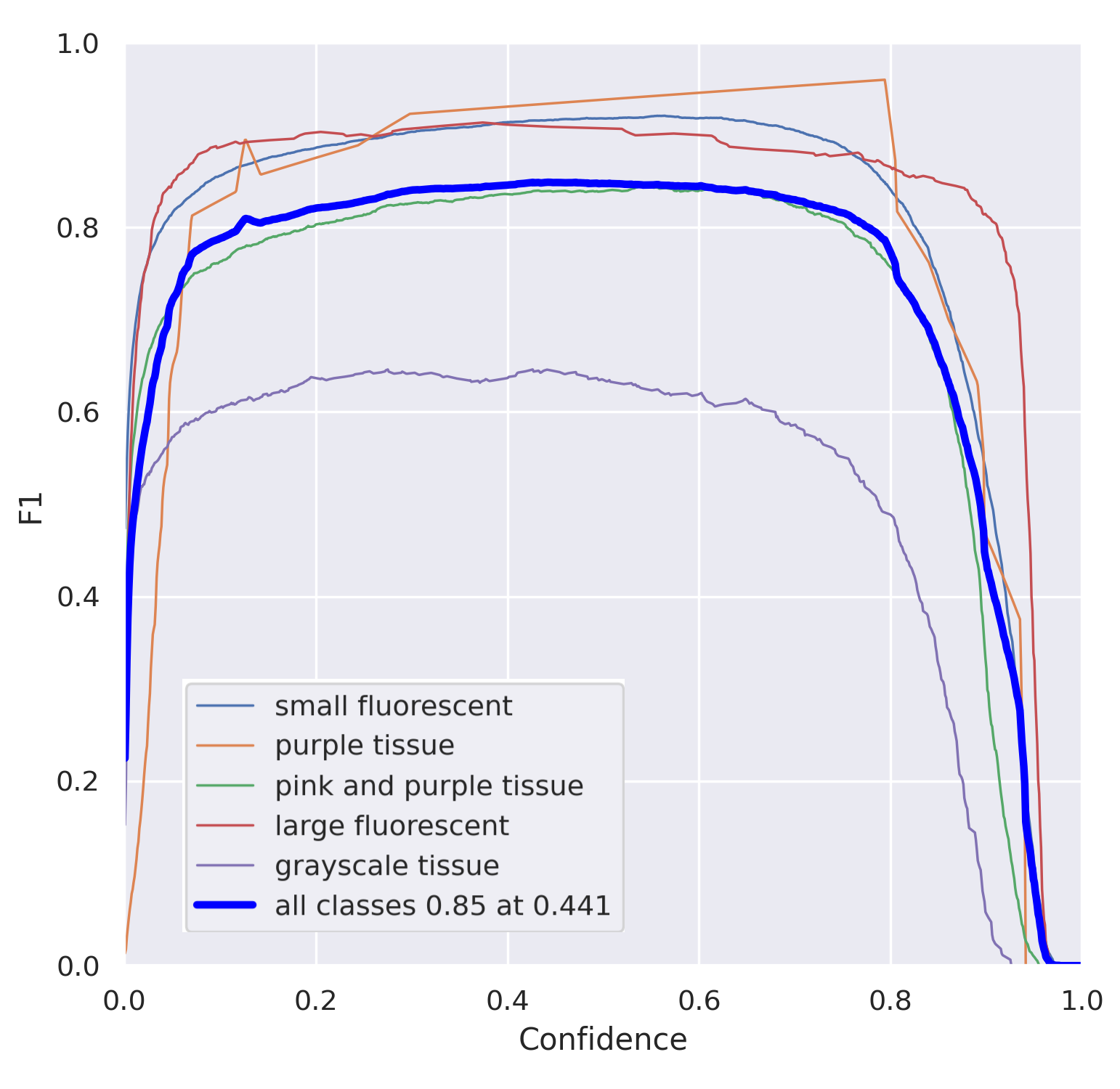}
        \label{fig:binaryPR}
        \caption{}
      \end{subfigure}
      \begin{subfigure}[b]{0.45\textwidth}
        \includegraphics[height=8cm]{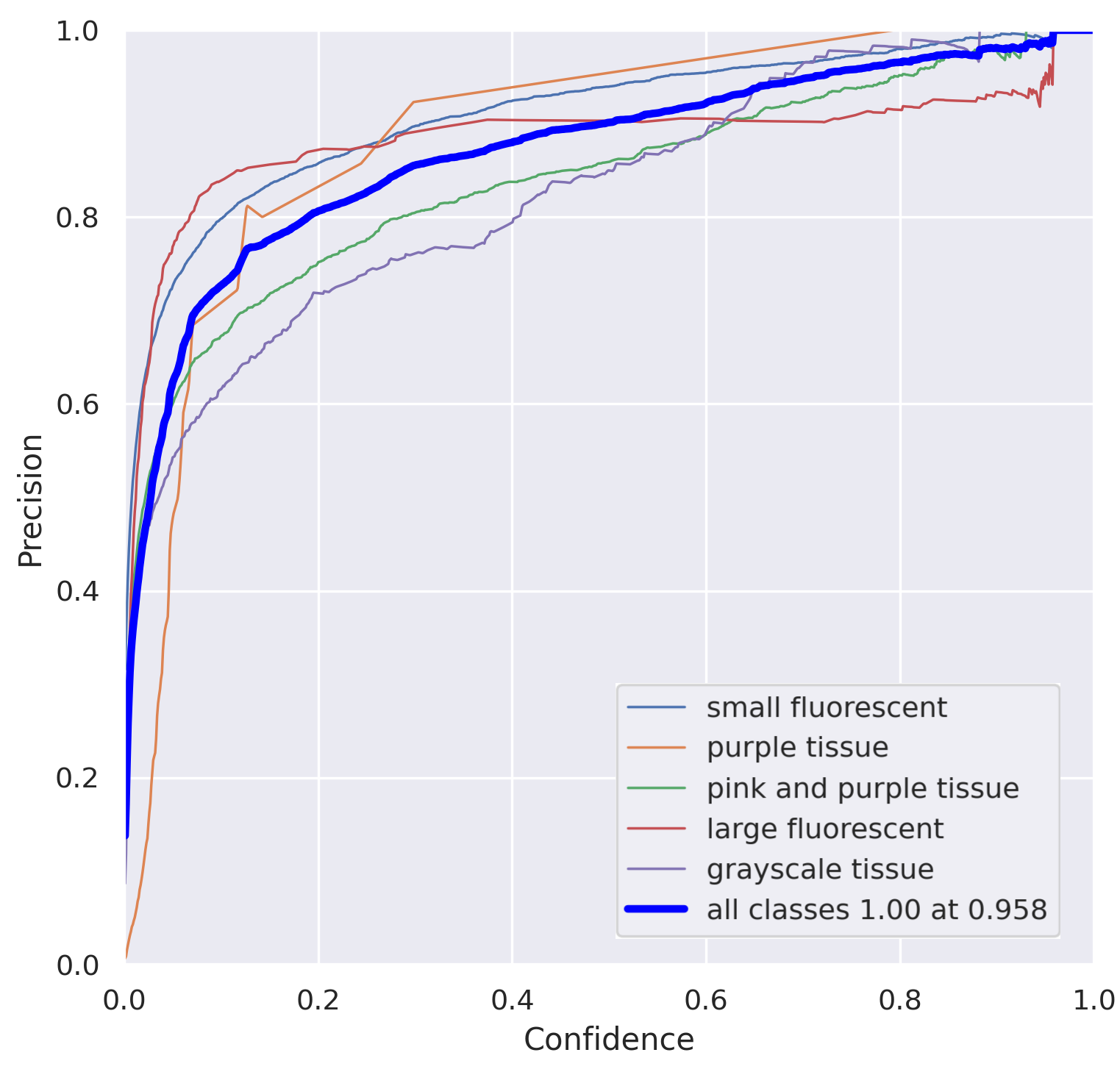}
        \label{fig:binaryPR}
        \caption{}
      \end{subfigure}  
    
      \begin{subfigure}[b]{0.45\textwidth}
        \includegraphics[height=7.8cm]{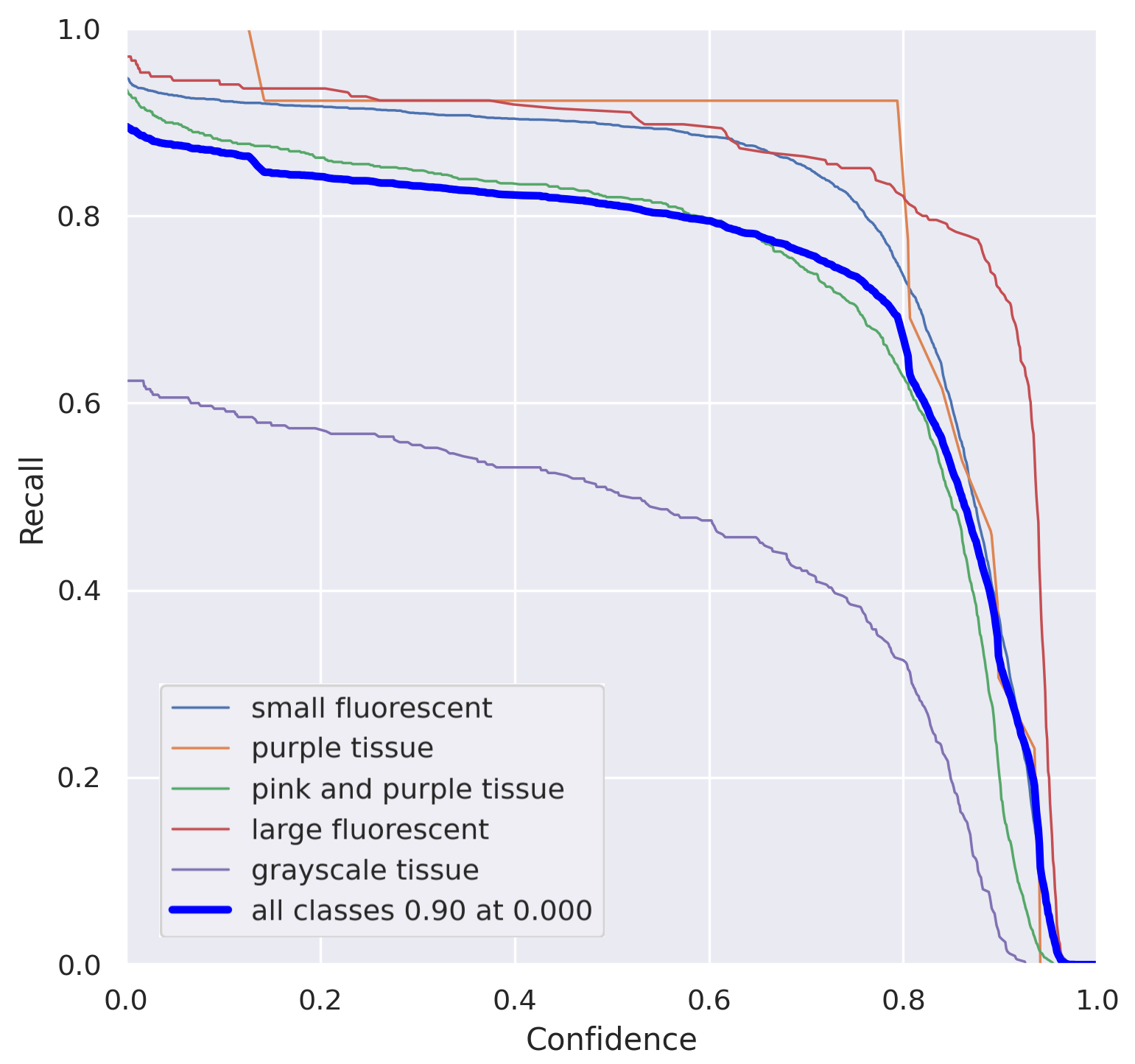}
        \label{fig:radarPlot}
        \caption{}
      \end{subfigure}
      \begin{subfigure}[b]{0.45\textwidth}
        \includegraphics[height=8.2cm]{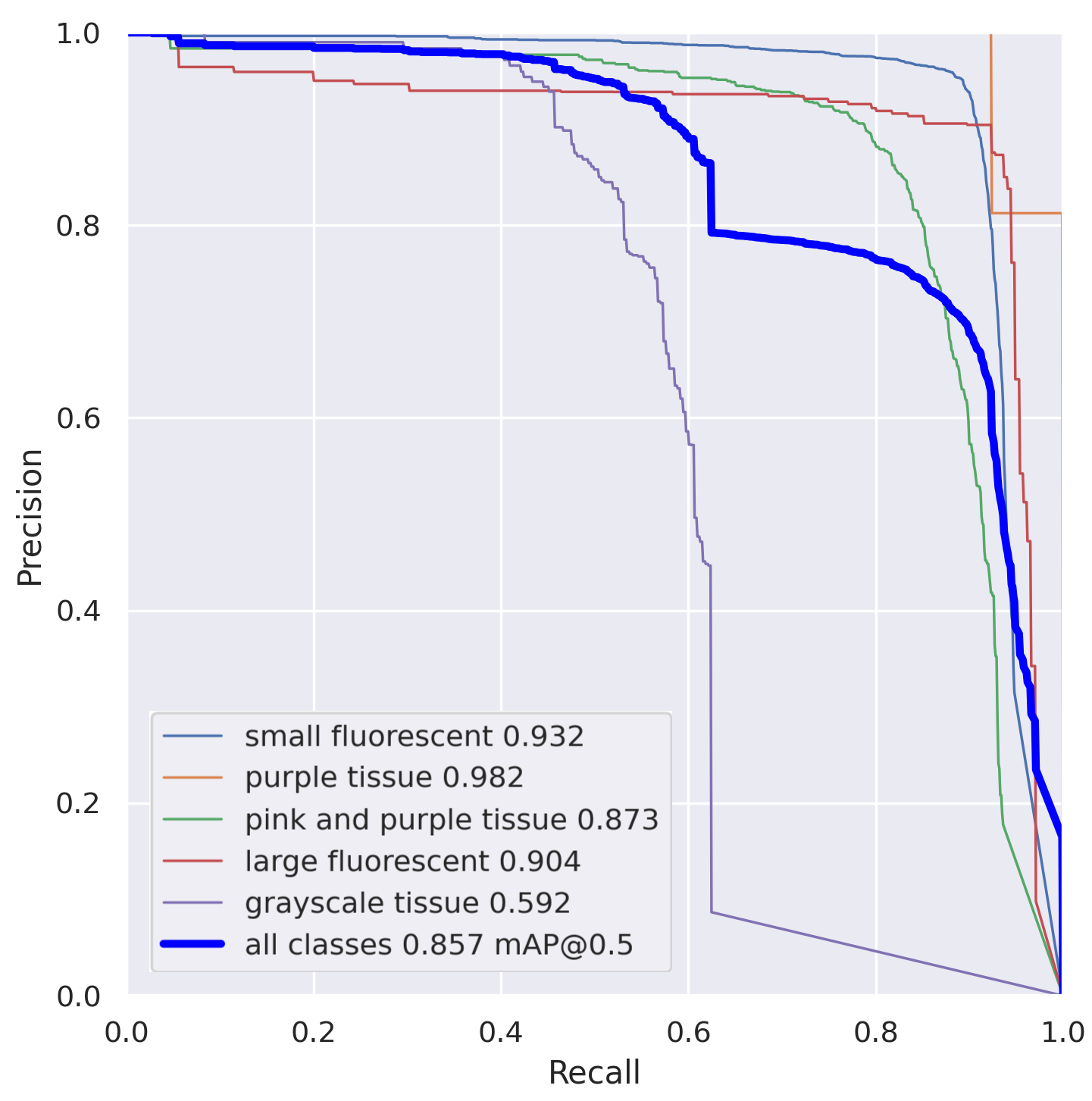}
        \label{fig:radarPlot2}
        \caption{}
      \end{subfigure}
      \caption{Experimental evaluation based on (a) F1 score per class, (b) Precision per class, (c) Recall per class, and (d) Precision-Recall per class. The comparatively bad performance for grayscale tissue samples is a results of unbalanced data in the competition training set. Only purple tissue samples occur less frequent in the training dataset than the grayscale tissue samples, and yet their metrics are significantly better. Intuitively, the segmentation complexity for purple tissue samples is lower due to a clear boundary between cell and background.  Best viewed in color.}
    \end{figure*}

    %\begin{figure*}
    %  \centering
    %  \begin{subfigure}[b]{0.24\textwidth}
    %    \includegraphics[height=4cm]{images/mutation1.png}    
    %    \label{fig:binaryPR}
    %    \caption{}
    %  \end{subfigure}
    %  \begin{subfigure}[b]{0.24\textwidth}
    %    \includegraphics[height=4cm]{images/mutation2.png}    
    %    \label{fig:binaryPR}
    %    \caption{}
    %  \end{subfigure}
    %  \begin{subfigure}[b]{0.24\textwidth}
    %    \includegraphics[height=4cm]{images/mutation3.png}    
    %    \label{fig:binaryPR}
    %    \caption{}
    %  \end{subfigure}
    %  \begin{subfigure}[b]{0.24\textwidth}
    %    \includegraphics[height=4cm]{images/mutation4.png}    
    %    \label{fig:binaryPR}
    %    \caption{}
    %  \end{subfigure}
    %  \caption{Graph mutations that optimize node positions and distribution to achieve best possible segmentation. Mutation steps are color-coded from blue to yellow for later mutation steps. Remarkably the mutations predominantly adapt for magnifications or relative node distance to the original image while remaining a structure that features nearly equidistant nodes level-wise. Best viewed in color. }
    %\end{figure*}

\subsection{Evaluation}
The experimental evaluation and comparison with best-performing methodologies from the 2018 data science bowl \cite{DataScienceBowl} can be found in Table \ref{tab:mainComparison}. The F1 score is closely correlated to the competition score. For various IoU thresholds between target and predicted segmentation F1 scores were computed and averaged. Our approach is best in terms of competition score, average F1 score, and recall at 0.7 IoU. While our approach performs best with the least pixel-wise missed at 0.7 IoU, in terms of extra predicted pixels, our approach just scores second after the Mask-RCNN approach \cite{MaskRCNN} and shortly followed by the approach of \cite{Unet}. 

In terms of extra and missing prediction, pixel-wise alignment compared to the annotation is visualized in the comparative segmentation maps (Figure \ref{fig:segmentationMaps}). Red areas show extra prediction of cell pixels, where missing prediction is shown in blue. In such a representation an ideal segmentation would reduce the blue and red areas to zero. For medical applications, false positive error (type I) or extra annotation would be preferable over false-negative errors (type II) or missing annotation. While our approach generally performs better than the compared approaches with less area for missing and extra pixels the difference in performance for different experimental variations is still significant. 

The ability to generalize well for various cell types can be evaluated when the comparative F1 score at 0.7 IoU is calculated for each cell class. In the official dataset, a categorization of cell types is not provided, such that the knowledge of cell types cannot be used to improve model performance in the training stage. For that reason, manual classification for the test set was performed. The per-cell F1 scores at 0.7 IoU are presented in Table \ref{tab:ExperimentalVariation} and visualized in Figure \ref{radarPlot}. In this form of visualization, a model with the best ability to generalize will result in covering a larger area than other model plots. In practice, a model still may be disqualified for use due to poor performance for one of the cell-type images.

Another part of experimental evaluation concerns, the mutation of graph structure and the understanding of optimization of neurons that learn node transitions when the features of lower magnification change due to a changed node position. In essence, one could assume that the modification of a low magnification node position that changes the node feature itself might set previously learned features back, or even that two different optimization goals are to be achieved. The optimization of graph structure, and the optimization of learned features in node transitions to segment cell nuclei. The node feature vector holds pixel information and position information in different dimensions. %We, therefore, observe best training results with

\section{Conclusions}
We proposed an algorithm that generalizes well in the task of nuclei segmentation of unknown histology images. We showed that the multi-magnification graph representation of an image allows for the application of graph neural networks that focus on message passing between nodes symbolizing pixels in multiple magnifications. In this way, the model capacity limitation of classical convolutional-based networks can be overcome by focussing on node transition instead of learning kernel patterns. The modification of graph structure in training allows to identify the ideal graph representation, thus extracting the best distances to learn node features relevant for accurate segmentation. Experimental evaluation showed superior performance to state-of-the-art segmentation algorithms.

%\clearpage

\section*{Authors}

\textbf{Yoav Alon} received a Master of Science in Computer Science at the University of Leicester and is currently a PhD-Student at the School of Informatics of the University of Leicester. %Reserach in Reinforcement Learning, Computer Vision, Robotics and Medical Image Analysis. 
 
\textbf{Huiyu Zhou} received a Bachelor of Engineering degree in Radio Technology from Huazhong University of Science and Technology of China (1990), and a Master of Science degree in Biomedical Engineering from University of Dundee of United Kingdom (2002), respectively. He was awarded a Doctor of Philosophy degree in Computer Vision from Heriot-Watt University, Edinburgh, United Kingdom (2006). Dr. Zhou currently is a full Professor at School of Computing and Mathematical Sciences, University of Leicester, United Kingdom. He has published over 380 peer-reviewed papers in the field. He was the recipient of "CVIU 2012 Most Cited Paper Award", “MIUA 2020 Best Paper Award”, “ICPRAM 2016 Best Paper Award” and was nominated for “ICPRAM 2017 Best Student Paper Award” and "MBEC 2006 Nightingale Prize". His research work has been or is being supported by UK EPSRC, MRC, EU, Royal Society, Leverhulme Trust, Puffin Trust, Alzheimer’s Research UK, Invest NI and industry. 

\nocite{*}
%\newpage
%\clearpage

%%Harvard
\bibliographystyle{model2-names.bst}\biboptions{authoryear}
\bibliography{refs}

\end{document}